# "Tall wheatgrass (*Thinopyrum ponticum* (Podp)) in a real farm context, a sustainable perennial alternative to rye (*Secale cereale* L.) cultivation in marginal lands"

Carlos S. Ciria[a,b]*, Carlos M. Sastre[a], Juan Carrasco[a], and Pilar Ciria[a]

[a] *CEDER-CIEMAT – Centro de Desarrollo de Energías Renovables, Autovía de Navarra A-15, Salida 56, 42290 - Lubia (Soria)*

[b] *Departamento de Producción Agraria. Universidad Politécnica de Madrid (UPM), 28040 Madrid, Spain*

\* *Correspondence:* carlossixto.ciria@ciemat.es; Tel.: +34-975-281-013 (Carlos S. Ciria)

**ABSTRACT**: In order to face the expected increasing demand of energy crops without creating conflicts of land occupation sustainability, farmers need to find reliable alternatives in marginal agricultural areas where the production of food hardly ever is economically and environmentally sustainable. The purpose of this work was the study of the viability of the introduction of new non-food crops in marginal areas of real farms. This study compares the profit margin and the energy and environmental performance of growing tall wheatgrass, in the marginal area of a rainfed farm versus rye, the annual crop sowed traditionally in the marginal area of the farm. The cited farm owned 300 ha of which about 13% was marginal. The methodology was based on the use of the profit margin of the crops as indicator for the economic assessment and Life Cycle Assessment (LCA) as technique for the energy and the environmental evaluations. Results of the economic analysis showed a slight enhancement of the profit margin for tall wheatgrass (156 €·ha$^{-1}$·y$^{-1}$) compared to rye (145 €·ha$^{-1}$·y$^{-1}$). Environmental LCA was driven by $CO_2$ fixation due to soil organic matter increase and reduced inputs consumption for tall wheatgrass that produced a Global Warming Potential (GWP) of – 1.9 Mg $CO_2$ eq.·ha$^{-1}$·y$^{-1}$ versus 1.6 Mg $CO_2$ eq.·ha$^{-1}$·y$^{-1}$ obtained for rye. Tall wheatgrass cultivation primary energy consumption was less than 40% of rye's consumption. According to the results achieved it was concluded that tall wheatgrass is better option than rye from the energy and the environmental point of views and slight better option from the economic view. Considering these results, monetarization of the $CO_2$ eq. reductions of tall wheatgrass compared to rye is essential to improve its profit margin and promote the implantation of this new crop in marginal areas of farms.

## 1. Introduction

Nowadays, agriculture has multiple challenges to address, due to the continuous rise of the demand for food and fiber with more efficient and sustainable production methods (FAO, 2009). Another challenge of the modern farming is the depopulation of rural areas derived from the constant job losses in the primary sector due to the economy modernization (Deggans et al., 2019) and the progress of the automation (Udell et al., 2019). In this context, the cultivation of dedicated energy crops on utilized arable productive lands increasingly creates a conflict with food production and poses important issues on the sustainability of this option (Bordonal et al., 2018). Example of them is the support given to corn (*Zea mays* L.) cultivation for biogas production in countries of Central Europe, which is causing severe consequences to the environment (Mardoyan and Braun, 2015), when there were other production methods more environmental sustainable (Marousek, 2013). However, new techniques for enhancing the yields of energy crops (Marousek, 2014) and the use of perennials in marginal lands as a potential source for biomass production (Shujiang et al., 2013) are seen as sustainable ways to overcome previous problems., although the poor profit margin has been identified as one of the main barriers to be broken down (Liu et al., 2011).

The last European Common Agriculture Policy reform (CAP) approved in 2015 has introduced payments linked to environmental and climate good practices (*greening practices*) committing around 50% of the direct total payments to farmers who have to introduce changes in their crop rotations. In this respect, perennial grasses may represent a new alternative that can be integrated in farming intensification schemes as fodder and livestock production systems, combining high yields with economic and environmental benefits (Sulas et al., 2015) and allow farmers to meet CAP requirements in degraded areas. In considering potential new crop alternatives, to traditional crops, the use of those grasses has been widely reported as an environmentally sustainable option in the EU (James, 2010). Historically, rye has been the traditional cereal established in the lower cereal yielding areas, namely marginal lands, due to its well-known high rusticity and better adaptation compared to wheat (Wrigley and Batey, 2010).

Many studies in literature provide agronomic (Sastre et al., 2016), economic (Ciria et al., 2019) and environmental (Sastre et al., 2014) data on energy crops in marginal agricultural lands. The studies are based on the results obtained in isolated experimental and demonstration parcels. Also, there are studies at big scale that only are focused on the agronomical part of the energy crops (Gominho et al., 2011) but very scarce if any information is available on the impact that the introduction of the new crops in marginal areas of real farming systems may have on the economic balance and environmental management of the farms.

In the above context, the main aim of this work is to evaluate if tall wheatgrass is better option than rye considering the economic, environmental and energy performance of both species cultivated in the marginal lands of a real rain-fed farm. Thus, producers can have new scientific evidences for making a decision about what alternative is better to sow in marginal areas. The economic performance of traditional crops grown in the non-marginal land area of the farm has also been assessed to construct the economic figures of the farm and to evidence the differences with respect to marginal land crops profitability in the context of a single farm.

## 2 Materials and methods

### 2.1 Study farm description

The farm objective of study is a representative farm of the North-Central Spanish Plateau, one of the most extended winter cereals farming areas in South Europe, but strongly affected by pedoclimatic and socioeconomic marginal factors. It is defined by Joint Research Centre (JRC) as a cropland with medium-low intensity and severe limitations, with low productivity rates (Meyer, 2014). The farm is dedicated to winter cereals cultivation under rain-fed conditions with traditional tillage practices mainly.

The farm had an extension of 302 ha distributed in 38 plots with an average area of 7.9 ha (maximum surface was 37 ha and minimum 1 ha per plot). The maximum distance between farm fields was 15 km. A total of 40 hectares of the farm surface were considered marginal land due to the economic constraints (low cereal yielding) and biophysical constraints (low organic matter content (OM) (≤1%) and stoniness (≥15% in soil volume).

The farm had its own machinery required for crop mechanization and had 2000 $m^2$ facilities for grain storage and agricultural equipment guard.

### 2.2 Farm management

The main purpose of the farm was the production of grain and the byproduct of this activity was the straw that was also sold. Conventional tillage was used in farm production systems. The species sown out of the marginal land area during the four years studied were principally winter cereals and also sunflower (*Helianthus annus* L.). The winter cereals were wheat (*Triticum aestivum* L.) that was the main crop (111 ha of land occupation on average), barley (*Hordeum vulgare* L.) with mean annual surface of 62 ha, triticale (*Triticosecale x* Wittmack) with annual average distribution in 32 ha. Sunflower annual distribution averaged 30 ha and fallow land 27 ha, in the study period. Commercial seeds multiplied by the previous years in the farm were used by the farmer, except sunflower hybrid seeds that were purchased each year. The traditional crop in the marginal areas of the farm was rye.

The methodology to sow cereals in the farm consisted of two tillage labors and one fertilizer application in autumn (early October to December) of NPK complex before sowing. After sowing, nitrogen fertilization with calcium ammonium nitrate with 27% of richness in nitrogen (CAN 27%), roller pass in early spring (mid-March) and phytosanitary application for weed control in April and early May to barley, triticale and rye. Herbicide treatment was done during late autumn for wheat. Sowing period for cereals went from October to early December.

Sunflower was sown in late May and harvested in mid-October. Fields were ploughed in early winter and then, cultivator was passed twice before sowing in spring. Herbicide glyphosate was applied before sowing.

Fallow land was ploughed in late spring and cultivated in summer for preparing the fields for the next sowing period.

Tall wheatgrass fields were established on marginal land area during early autumn in 2013. Land preparation was similar to winter cereals with the same fertilization. A phytosanitary treatment was applied in the establishment year for dicotyledonous weeds. Top dressing fertilization in spring and crop mowing and baling in the first days of August were the unique recurrent field works performed every year.

Table 1 shows the inputs used in the farm crops during the four years of the study.

**Table 1**: Farm inputs and doses used for each crop during the study years. All data are on an annual basis, except when otherwise indicated.

|  | Non marginal land | | | | Marginal land | |
| --- | --- | --- | --- | --- | --- | --- |
|  | Wheat | Triticale | Barley | Sunflower | Rye | Tall wheatgrass |
| Cultivar | Berdun | Verato | Cometa | Krisol | Petkus | Szarvasi-1 |
| Sowing dose (Mg·ha$^{-1}$) | 0.20 | 0.20 | 0.18 | 0.004 | 0.15 | 0.02 |
| Product (NPK) | 8-24-8 | 8-24-8 | 8-15-15 | - | 8-24-8 | 8-24-8 |
| Base fertilization (Mg·ha$^{-1}$) | 0.30 | 0.25 | 0.30 | - | 0.20 | 0.30 (Establish year) |
| Product | Calcium ammonium nitrate 27% | Calcium ammonium nitrate 27% | Calcium ammonium nitrate 27% | - | Calcium ammonium nitrate 27% | Calcium ammonium nitrate 27% |
| Top fertilization (Mg·ha$^{-1}$) | 0.30 | 0.20 | 0.20 | - | 0.15 | 0.15 |
| Herbicide | Clorsulfuron 75% + Clortoluron 50% | 2.4-D acid 60% | Metsulfuron methyl 11,1% Tribenuron-methyl 22,2% + 2,4-D acid 34,5% + MCPA 34,5% (2-methyl-4-chlorophenoxy acetic acid) | Glyphosate 45% | 2.4-D acid 60% | 2.4-D acid 60% |
| Herbicide dose (units·ha$^{-1}$) | 20 g +2 L | 0.8 L | 45 g +1 L | 2.5 L Presowing | 0.8 L | 1 L |

Cereals were harvested in mid-July; sunflower in mid-October and the cereal straw was baled after finishing the harvesting period. Tall wheatgrass was mowed in early August leaving the forage 4-5 days on the field for tedding and finally baled.

*2.3 Pedoclimatic conditions*

The climate in the study area is Continental Mediterranean characterized by cold winters, warm summers and low rainfall level. Meteorological conditions in the farm area corresponding to the study period were taken from Spanish Meteorological Agency (AEMET) nearer station. The mean annual temperature during study was 11.0ºC, with a maximum absolute temperature on July 2015 of 36ºC and a minimum absolute temperature on January 2017 of -11.7ºC. The total average annual rainfall was 513 mm being the wettest year 2014 (595 mm) and the driest 2017 (315 mm). The extreme free frost period during the study occurred from early May to mid-September. Drought period was usually from July to August.

Farm marginal land fields were characterized by poor sandy soil (88% sand, 8% lime, and 4% clay), 0.54 % organic matter content, 0.03% of nitrogen content, 6.73 mg·kg$^{-1}$ of absorbable phosphorus and 58.19 mg·kg$^{-1}$ of potassium, with a bulk density of 1.37 Mg·m$^{-3}$ and a content in coarse elements (> 2mm) of 29.58% in volume. Marginal soil showed high drainage, in consequence, low water retention, low nutrient content, high losses by leaching and no salinity problems. In general, it is a low productivity soil for most common food crops.

Soil samples were taken from the upper layer (0-30 cm) according to ISO 10381-1; texture (ISO 11277), pH (ISO 10390), electrical conductivity (ISO 11265), content in OM (ISO 10694), Phosphorus Olsen content (ISO 11263), potassium content (ISO 11260) and ammonium and nitrates content (ISO/TS 14256-1 EX.).

*2.4 Marginal land management*

In order to study the most suitable alternative for the marginal land of the farm, two experimental plots of rye (the traditional food crop in this type of land) and tall wheatgrass, as an alternative crop, of two hectares each were established in the marginal area of the farm in October 2013 managed as it was comment in section 2.2. Economic, energetic and environmental analysis of both parcels will be made in the four years of study (October 2013 to September 2017).

*2.5 Economic analyses*

Farm profit margin analyses, including the two alternatives on the marginal area described in section 2.4 was performed taking into account the prices of commodities (seeds, fertilizers, herbicides, fuels) and machinery used by the farmer during the study period (4 years). The labor and machinery costs including farming implements for each agricultural activity were calculated according to the technical characteristics, the real capacity for carrying out their work, the utilization costs and the conditions of the farm such as soil type. The calculation of them were carried out using the online database provided by Spanish Agriculture Department (MAPA, 2018). The differences between the costs and the total sales plus CAP aid were the results showed in the economic balance.

The economic study of the non marginal part of the farm was carried out calculating the weighted average of inputs and outputs of the different crops sowed in the farm during the study years with the real price before taxes of the invoices provided by the farmer. In the marginal area the costs calculations were performed in the same way by extrapolating the results of the experimental plots (2 ha each) to all marginal land area (40 ha) of the farm, due to this area presented similar biophysical constraints and equivalent cereal production (rye) along the years.

Tall wheatgrass implantation costs were spread over the four years of the study, although there are experiences in the region of the study that reported competitive yields of that crop during six years (Ciria et al., 2017) at least and other studies for the same crop reported life spans from 10 to 15 years (Sandor, 2011).

Same price of cereal straw for energy or animal use according to the Spanish market prices during the study period were considered for tall wheat grass biomass, due to the current lack of market for this product.

*2.6 Environmental and energy assessment.*

2.6.1 Goal and scope

The objective of the assessment was to determine and compare the energy consumptions and the environmental impacts of the two crop alternatives grown in the marginal area of the farm described in section 2.2, from a life cycle perspective. LCA is the environmental management tool, regulated by ISO 14040 and ISO 14044, selected to perform the evaluations.

The type of LCA conducted was attributional and from cradle to farm gate (see Fig. 1). Limits of the study included: a) the production and transport of agricultural inputs consumed; b) the manufacture and use of machinery including diesel consumption as well as exhaust emissions affecting to the Global Warming Potential (GWP), and c) fertilizer application and associated field emissions affecting to GWP. Simapro 9 software tool and Ecoinvent 3.5 European database were used to conduct the LCAs. The "cut-off /allocation, recycled content" version of this database was selected for the modeling because it follows the same approach as Ecoinvent 1 and 2 and therefore allows better comparisons with other LCA studies that used this database.

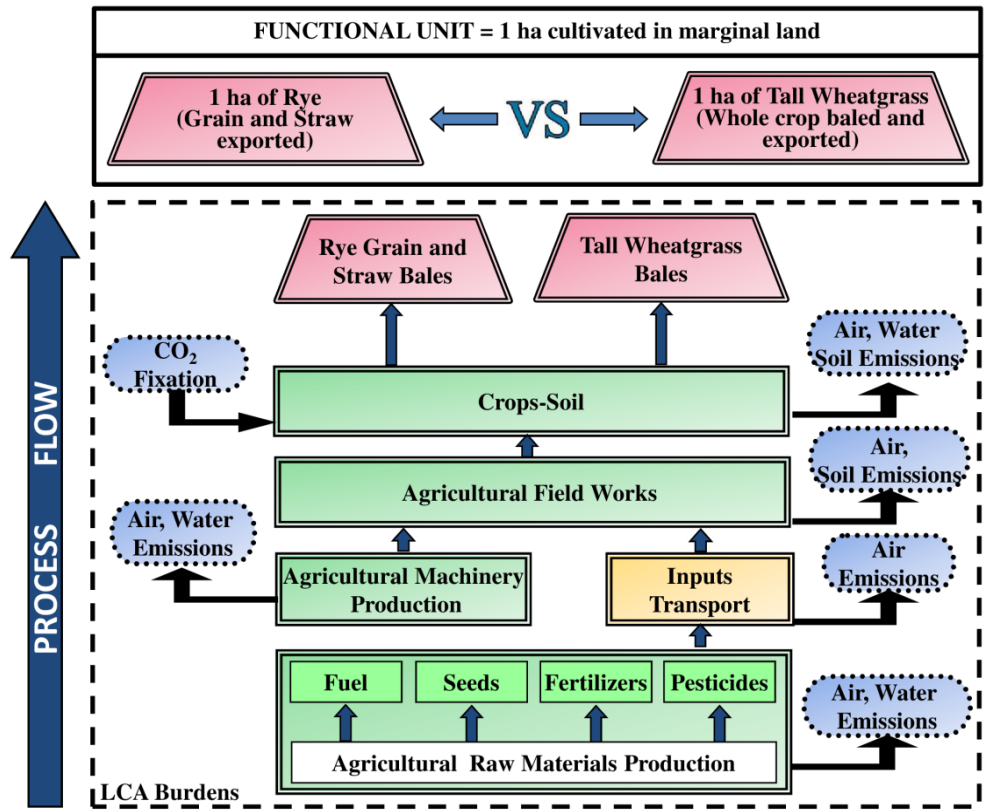

**Fig. 1**. Environmental and energy assessment graphical system description. It includes the flow diagram with all the processes considered, the limits of the system as well as the functional unit specified for the two cropping alternatives on the farm's marginal land area under study.

As shown in Fig. 1, the functional unit selected for the assessment is 1 ha of each of the two alternatives tested in the marginal area of the farm.

### 2.6.2 Life Cycle Inventory analysis

The methods used for the inventory modeling followed those described in Sastre et al. (2016) that affected GWP and energy calculations and were relevant for crop cultivation phase. The modeling of the inventory was made coherent with the economic analyses. All the data needed as input for the previous methods and the processes for its obtaining are described in detail in this section. Annual field works and inputs considered were the same every year for rye and for tall wheatgrass the extra inputs and fieldworks of the implantation year were spread by crop life span (section 2.5) to obtain a representative average. Crops average productivity, fertilizer, herbicides and sowing seeds consumption were taken from Table 1. Diesel consumption of agricultural machinery as well as its amortization (proportion of production and maintenance attributable to its use for crop cultivation) were calculated according to the fieldworks performed (section 2.5) using the Database of Spanish Agriculture Department (MAPA, 2018). Background processes for modeling all the inventories were taken from Ecoinvent 3.5.

The sowing seed production was modeled considering a seed production yield equal to grain yield for rye (Table 2) and of 0.165 Mg·ha$^{-1}$·y$^{-1}$ for tall wheatgrass. Tall wheatgrass seed yield was estimated as 3% of the crop biomass yield (Table 2) (Liu and Wang, 2011). Inputs considered for sowing seed inventory modeling where the same as the ones consumed for the alternatives evaluated plus additional transport and seed processing energy consumption as described in (Nemecek et al., 2007). Seed sowing doses were 0.02 Mg·ha$^{-1}$ for tall wheatgrass and 0.15 Mg·ha$^{-1}$ for rye.

Annual field works performed for tall wheatgrass cultivation resulted in diesel consumption of 31.95 L·ha$^{-1}$ and in amortization of machinery of 0.00099 Mg·ha$^{-1}$ of tractor, 0.00020 of Mg·ha$^{-1}$ of tillage implements and of 0.00251 Mg·ha$^{-1}$ of other implements on average. For rye, average diesel consumption was 55.39 L·ha$^{-1}$ and average amortization of machinery was 0.00135 Mg·ha$^{-1}$ of tractor, 0.00098 Mg·ha$^{-1}$ of harvester, 0.00082 of Mg·ha$^{-1}$ of tillage implements and of 0.00201 Mg·ha$^{-1}$ of other implements.

Agricultural machinery exhaust emissions were considered according to diesel consumption (Nemecek et al., 2007). N$_2$O emissions (Nemecek et al., 2012) due to fertilizer application, crop residues decomposition and NH$_3$ conversion into N$_2$O were also accounted. Average annual N$_2$O emissions were of 0.000817 Mg·ha$^{-1}$·y$^{-1}$ for tall wheatgrass and of 0.001757 Mg·ha$^{-1}$·y$^{-1}$ for rye.

As conventional tillage does not increase organic matter (Balesdent et al., 2000) and cultivation of rye with straw exported is the common practice for the marginal land of the farm, it have been considered that the organic matter content (0.540 %) and the organic carbon content (0.313%) obtained from initial soil analyses (0-30 cm) were stable and corresponded to a point of equilibrium. In the literature there are studies that argue that rye enhances the soil organic matter only when it is used as a cover crop under conventional tillage conditions (Jarecki and Lal, 2003) but there is no evidence found suggesting that rye enhance the soil organic matter when conventional tillage is carried out and grain and straw are harvested and exported. Therefore, the alternative that considered rye cultivation was a continuation of previously described common practices in the marginal land of the farm and therefore this did not change the soil composition equilibrium. Due to this fact, no CO$_2$ credit or debt was considered because of changes in soil organic matter and organic carbon for rye cultivation system. Nevertheless, in the case of tall wheatgrass due to its perennial character, the accumulation of belowground biomass and the absence of tillage operations (Signor et al., 2018), an increment of soil organic matter and carbon was expected. This absence of tillage also contributes to maintain organic matter quality due to the lower activity of soil microorganism that will help to keep it more stable (Sanchez-Gonzalez et al., 2017). These increments were confirmed by soil analyses of the experimental parcels performed after three year of crop implantation that revealed a raise up to 0.677 % and 0.393% of the soil organic matter and organic carbon contents, respectively. The increase of soil organic carbon corresponded to an annual fixation of 0.765 Mg C·ha$^{-1}$ equivalent to 2.805 Mg CO$_2$·ha$^{-1}$ These figures were obtained by using soil bulk density (1.370 Mg soil·m$^{-3}$) of the 0-30 cm layer and considering the influence of coarse elements (>2 mm) (29.58% in volume) that were previously removed from the samples used to determine organic matter contents.

### 2.6.3 Impact assessment methods

Two impact assessment methods were selected to transform Life Cycle Inventory (LCI) elements into impacts.

The impact assessment method for the evaluation of the GWP was the Intergovernmental Panel on Climate Change (IPCC) 2013 for '100 years' time horizon (Frischknecht et al., 2010).

Cumulative Energy Requirement Analysis (CERA) was the method selected to assess the primary energy consumed by the two alternatives evaluated (Frischknecht et al., 2010).

## 3 Results and discussion

### 3.1 Farm profit balance

As expected, the most profitable crops were the crops sowed in the non marginal area of the farm (Table 2): wheat: 416.62 €·ha$^{-1}$·on weighed annual average during the studied period, barley: 339.31 €·ha$^{-1}$·y$^{-1}$ and sunflower: 316.04 €·ha$^{-1}$·y$^{-1}$. The figures included the CAP aids, for a total of 165 €·ha$^{-1}$·y$^{-1}$ in all cases.

The crops in the marginal land area also achieved positive results although they had much lower profit margin than those in the non-marginal zones of the farm, being slightly more profitable the option of sowing tall wheatgrass (156.19 €·ha$^{-1}$·y$^{-1}$) compared to the rye alternative (145.14 €·ha$^{-1}$·y$^{-1}$). However, the profit was negative for both crops when CAP aids were not considered (Table 2) thus showing that those payments were essential to obtain a positive profit margin in the case of crops in the marginal area. In fact, without CAP subsidies the economic balance was at -19.86 €·ha$^{-1}$·y$^{-1}$ for rye and -8.81 €·ha$^{-1}$·y$^{-1}$ for tall wheatgrass. The main costs (Table 2) of marginal land crops were the ones associated to the machinery and labor (54% rye and 53% for tall wheatgrass), followed by the fertilizers costs (34% rye and 33% for tall wheatgrass). Seeds (10% rye and 14% for tall wheatgrass) and herbicides (1.8% rye and 0.7% for tall wheatgrass) were less relevant costs items.

**Table 2**: Economical balance of the crops sowed in the study farm (weighted average annual values in the 4 years of the study period).

| (€·ha$^{-1}$) | Non Marginal land | | | | | Marginal land | |
|---|---|---|---|---|---|---|---|
| | Wheat | Barley | Triticale | Sunflower | Fallow | Rye | Tall wheatgrass |
| Seed | 44.48 | 37.11 | 36.54 | 40.86 | | 31.00 | 35.00 |
| Herbicide | 21.89 | 20.50 | 5.50 | 14.51 | | 5.50 | 1.73 |
| Fertilizer | 174.94 | 168.78 | 133.26 | 0.00 | | 103.80 | 81.30 |
| Machinery + Labor | 186.20 | 182.99 | 177.10 | 160.92 | 59.47 | 164.50 | 131.85 |
| **Total Cost** | **427.51** | **409.38** | **352.40** | **216.28** | **59.47** | **304.80** | **249.87** |
| Yield (Mg·ha$^{-1}$) | 3.38 | 3.11 | 2.56 | 1.14 | | 1.50 | |
| Selling price grain (€·Mg$^{-1}$) | 174.78 | 161.52 | 155.43 | 322.76 | | 158.69 | |
| Straw yield (Mg·ha$^{-1}$) | 2.03 | 1.86 | 1.54 | 0.00 | | 1.07 | 5.50 |
| Selling price straw (€·Mg$^{-1}$) | 43.83 | 43.83 | 43.83 | | | 43.83 | 43.83 |
| **Total Sales** | **679.13** | **583.68** | **466.09** | **367.33** | | **284.93** | **241.07** |
| **CAP aid** | **165.00** | **165.00** | **165.00** | **165.00** | **165.00** | **165.00** | **165.00** |
| **Balance without CAP** | **251.62** | **174.31** | **113.69** | **151.04** | **-59.47** | **-19.86** | **-8.81** |
| **Balance with CAP** | **416.62** | **339.31** | **278.69** | **316.04** | **105.53** | **145.14** | **156.19** |

If the marginal land of the farm (40 ha) would be completely sowed of tall wheatgrass the global income of the farm, taking into account the average surface of each crop cited in section 2.2 and the economical balance (table 2), would be 94,778.44 €·yr$^{-1}$; compared to 94,336.16 €·yr$^{-1}$ when sowing rye. This appears as a very poor economic incentive for farmer to grow tall wheatgrass, due to the final increment of the economic balance is only 0.5% compared to rye, and also considering the inconveniences concerning permanent land occupation and market risks associated to the new crop. In order to study the influence of increase the marginal area of the farm a projection estimation up to 50% of marginal surface referred to the total area of the farm, revealed a slight increment (2.3%) of the economic difference in favor of tall wheatgrass compared to rye option. However, important advantages of tall wheatgrass compared to rye such as the lower environmental impacts and energy consumption are brought into discussion in the next section 3.2.

The consumer awareness should help to evidence the environmental benefits of tall wheatgrass in addition to other type of benefits such as, the creation of self-employed people (Graessley et al., 2019) working in a context of growing digitalization and permanent connection of the whole agricultural value chain (Ludbrook et al., 2019).

*3.2 Farm global warming potential and primary energy evaluations for the marginal lands crop alternatives*

Results of GWP and primary energy are given by each alternative under study in Fig. 2, Fig. 3 and Fig. 4. Environmental and energy life cycle assessment results of tall wheatgrass were made with whole crop harvest and baled and results of rye were made grown to harvest the grain and bale the straw. In Fig. 2 the GWP for a 100 year time horizon was evaluated by phases using the IPCC 2013 methodology, for the tall wheatgrass alternative and rye alternative.

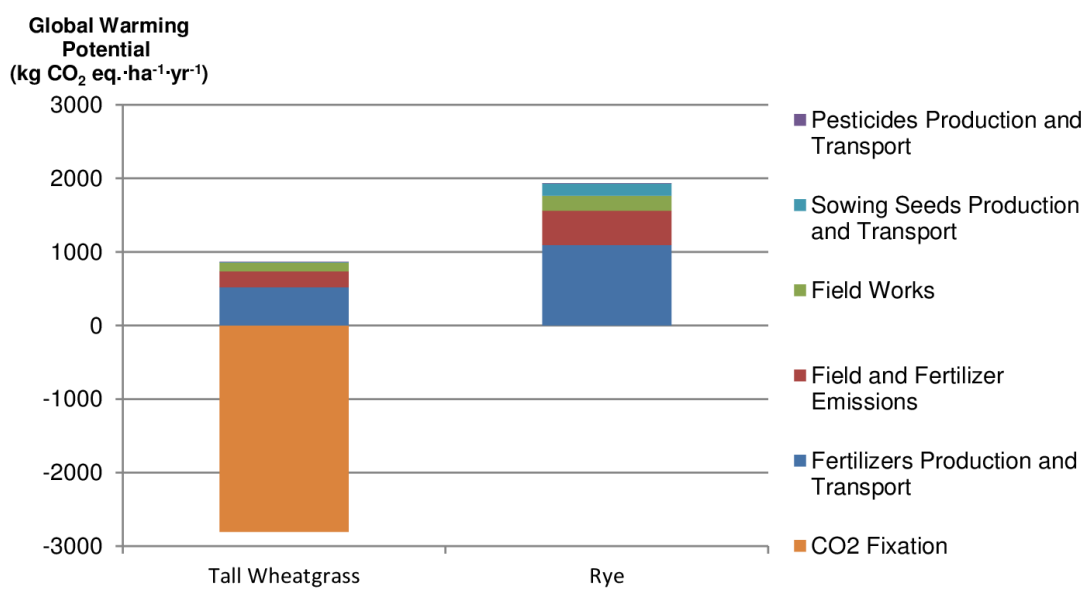

**Fig. 2**: Global Warming Potential for each alternative

Fertilizer production and transport was the phase that generated the highest impacts on the GWP accounted as the sum of all phases that generated positive impacts (Fig. 2), with 60.2% for tall wheatgrass and a 56.6% for rye. Field and fertilizers emissions derived from $N_2O$ release mainly, produced the second highest positive impact on GWP with 25.1% for tall wheatgrass and 24.1% for rye when straw is exported. Field works and seed production and transport had lower influence on GWP accounting, together for 14.6% for tall wheatgrass, and 19.2% for rye when straw is exported. Seed production produced lower impact on tall wheatgrass GWP emissions compared to rye (0.7% vs 8.7%) due to the smaller sowing dose required and the perennial character of that crop. Pesticides production and transport had the smallest effects on GWP impacts of all phases studied with figures below 0.2% of total GWP of the two alternatives.

All the phases of tall wheatgrass cultivation that contributed to increase the GWP generated together 0.863 Mg $CO_2$ eq.·ha$^{-1}$·y$^{-1}$, remarkably lower than the impact of these phases for rye cultivation (1.934 Mg $CO_2$ eq.·ha$^{-1}$·y$^{-1}$) (see Fig. 2). $CO_2$ fixation due to the increase of soil organic matter had significant influence for tall wheatgrass allowing a reduction of 2.805 Mg

$CO_2$ eq.·ha$^{-1}$·y$^{-1}$ of GWP in the evaluated period. This was equivalent to 325% of the GWP produced by the sum of the phases producing an increase of crop GWP. No net $CO_2$ fixation was considered for rye as growing this crop for grain and straw production is the traditional practice and it has been considered that it does not alter the soil organic matter balance (see 2.6.2). Negative GWP balance obtained for tall wheatgrass (-1.942 Mg $CO_2$ eq.·ha$^{-1}$·y$^{-1}$) could have remarkable importance if this positive externality is monetarized in form of subsidy for farmer in the future.

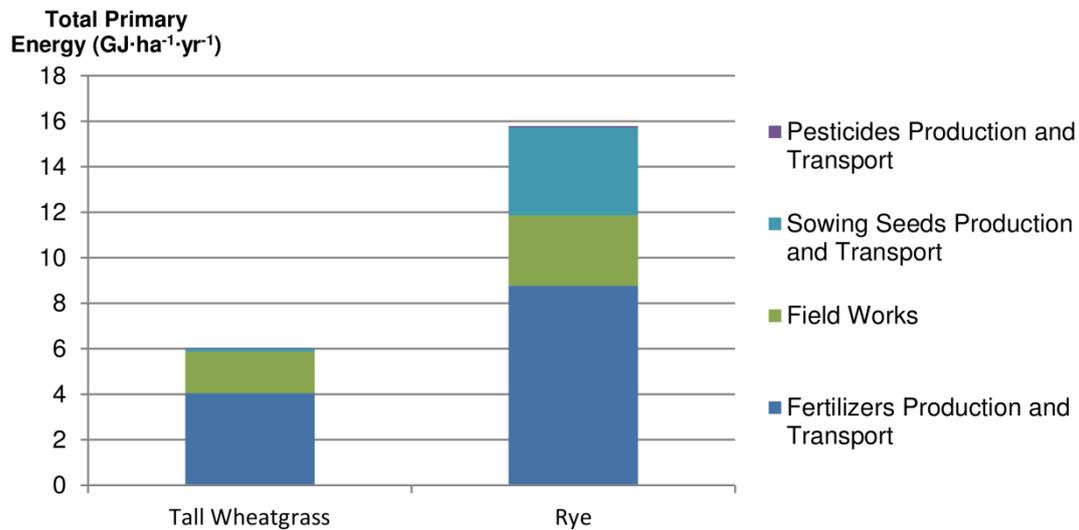

**Fig. 3**: Total Primary Energy consumed for growing tall wheatgrass and rye in the farm studied

Primary energy consumed for tall wheatgrass cultivation (6.0 GJ·ha$^{-1}$·y$^{-1}$) was approximately 38% of the energy consumed by rye with 15.8 GJ·ha$^{-1}$·y$^{-1}$ (Fig. 3). This difference between both crops is due to the perennial nature of tall wheatgrass because of sowing seeds, base fertilization and fieldworks for crop establishment are only required the first year of its life span. Fertilizer consumption was the most energy consuming phase for both alternatives under study with 67.1% for tall wheatgrass and 55.5% for rye, while sowing seed consumption was the second energy consuming phase for rye with 24.5% and the third one for tall wheatgrass with only 2.1% of total crop energy costs. Field works were the second energy consuming phase for tall wheatgrass with 30.4% and the third one for rye with 19.6%. Pesticides had very little influence in the primary energy consumption with less than 0.5% for the two crops evaluated.

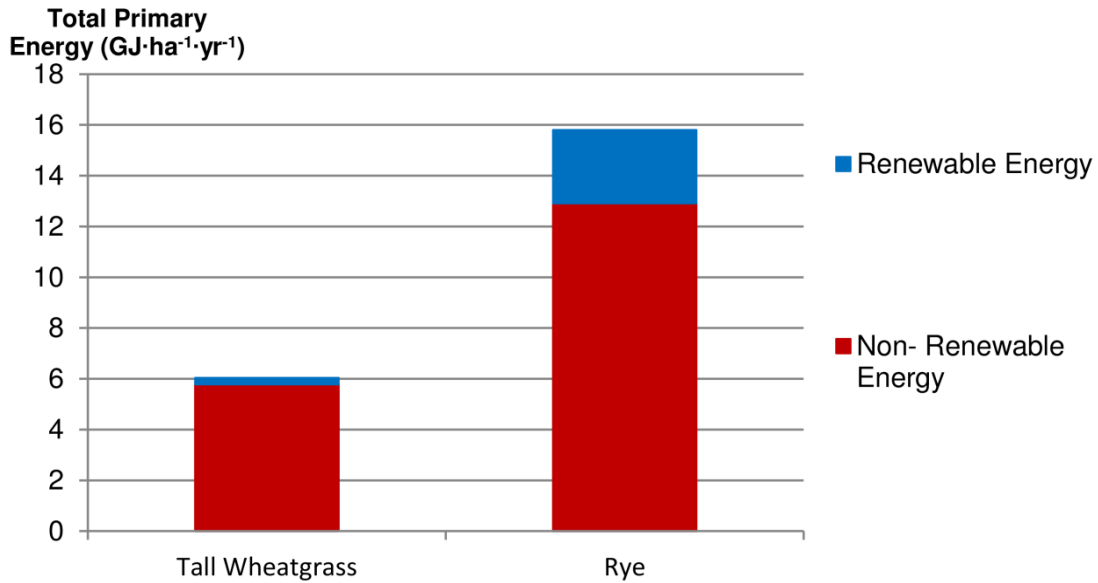

**Fig. 4**: Differences between renewable and non-renewable Primary Energy consumed for each alternative

Renewable primary energy consumption was much more important for rye (18.3%) than for tall wheatgrass (3.5%). This difference is mainly attributable to the energy content of the seeds utilized for sowing, given the much lower seed dose used for tall wheatgrass, 0.02 Mg·ha$^{-1}$ versus 0.15 Mg·ha$^{-1}$ for rye and the fact that seeds are only used the first year of crop establishment in the case of tall wheatgrass.. Non-renewable energy consumption, were the effects of sowing seed had less influence, was higher for rye with 12.9 GJ·ha$^{-1}$·y$^{-1}$ when compared to tall wheatgrass with 5.8 GJ·ha$^{-1}$·y$^{-1}$ on average in the studied period.

This study can be complemented in the future by testing other perennial grasses such as tall fescue (*Festuca arundinacea* Schreb.) or native grassland, different tillage techniques (Pearson et al., 2014), and new soil improvers (Marousek et al., 2019) to check the environmental and economic impacts in marginal areas of this wide range of alternatives.

This study can be replicated in more than 10.5 million ha of marginal land where tall wheatgrass can be cultivated only in Europe (Von Cossel et al., 2019). Moreover, according to FAOSTAT (FAO, 2020) between 2007 and 2017 there were more than 5 million ha sowed annually of rye in the world, located principally in Europe with an average yield of 2.04 Mg·ha$^{-1}$·yr$^{-1}$ that can be a potential sink of carbon transforming part of them in tall wheatgrass fields.

## 4 Conclusions

The profit margins obtained for tall wheatgrass and rye were negative for both crops grown in the marginal area of the studied farm when CAP aid was not accounted, but less losses were obtained for tall wheatgrass compared to rye. Tall wheatgrass produced a relevant annual increase of soil organic matter that together with the reduced fieldworks and inputs consumed for cultivation resulted in remarkable better carbon footprint for it (≈200% less), Moreover, the primary energy consumption for tall wheatgrass cultivation, was less than 40% of rye. According to these results tall wheatgrass is more sustainable than rye, but in spite of less crop management requirements for tall wheatgrass, the almost negligible difference between the profit margins of the two crops do not make attractive the option of tall wheatgrass against the traditional one, in particular considering the permanent land occupation and risks associated to the implementation of the new crop. Taking into account the monetarization of the positive externalities derived from soil carbon increment would be essential for the farmers to introduce this crop in marginal areas. Hence, the implantation of tall wheatgrass in marginal areas where traditionally are sowed by winter cereals will increase the carbon sequestration in the soil making the agriculture systems more sustainable compared to traditional management and achieving similar profit margin.

**Funding:** This research received funding from European Commission, in the frame work of HORIZON 2020 Programme, LCE-22-2016 topic call, within the project "Brazil-EU Cooperation for Development of Advanced Lignocellulosic Biofuels (BECOOL)" grant number 744821".

**Conflict of interest:** None.

**Data availability:** The datasets generated and/or analyzed during the current study are available from the corresponding author on reasonable request.